\documentclass[twocolumn,amsmath,amssymb,prl,superscriptaddress,showpacs,dvipdfmx]{revtex4}
\usepackage[dvips]{graphicx}
\usepackage{dcolumn}
\usepackage{bm}
\newcounter{one}
\usepackage{braket}
\usepackage[usenames]{color}
\definecolor{nblue}{rgb}{0.3,0.3,1.0}

\def\UTokyo{Department of Applied Physics, School of Engineering, The University of Tokyo,\\
7-3-1 Hongo, Bunkyo-ku, Tokyo 113-8656, Japan}
\def\UMainz{Institute of Physics, Johannes Gutenberg-Universit\"{a}t Mainz, Staudingerweg 7, 55128 Mainz, Germany}

\begin{document}

\title{Entanglement Swapping between Discrete and Continuous Variables}

\author{Shuntaro Takeda}
\affiliation{\UTokyo}
\author{Maria Fuwa}
\affiliation{\UTokyo}
\author{Peter van Loock}
\affiliation{\UMainz}
\author{Akira Furusawa}
\email{akiraf@ap.t.u-tokyo.ac.jp}
\affiliation{\UTokyo}

\date{\today}


\begin{abstract}
We experimentally realize ``hybrid'' entanglement swapping between discrete-variable (DV) and continuous-variable (CV) optical systems.
DV two-mode entanglement as obtainable from a single photon split at a beam splitter is robustly transferred by means of efficient CV entanglement and operations, using sources of squeezed light and homodyne detections. The DV entanglement after the swapping is verified without post-selection by the logarithmic negativity of up to 0.28$\pm$0.01. Furthermore, our analysis shows that the optimally transferred state can be post-selected into a highly entangled state that violates a Clauser-Horne-Shimony-Holt inequality by more than 4 standard deviations, and thus it may serve as a resource for quantum teleportation and quantum cryptography.
\end{abstract}



\maketitle

\begin{figure}[!b]
  \begin{center}
   \includegraphics[width=0.6\linewidth,clip]{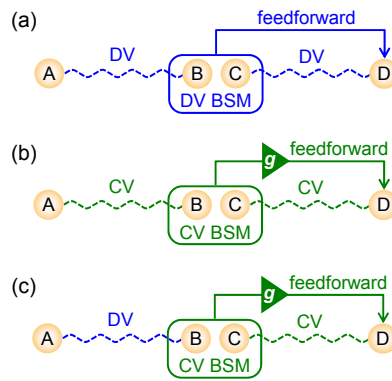}
  \end{center}
\caption{Schematic of entanglement swapping.
(a) DV entanglement swapping.
(b) CV entanglement swapping.
(c) Hybrid entanglement swapping.}
\label{NewFig1}
\vspace{-5mm}
\end{figure}

Quantum entanglement can be created between two distant quantum systems that have never directly interacted.
This effect, called entanglement swapping~\cite{93Zukowki,98Pan,02Sciarrino,05Riedmatten,99Tan,99vanLoock,04Jia,05Takei},
is a building block for quantum communication and computation~\cite{98Bose,01Knill,98Briegel,01Duan}.
It was originally proposed and demonstrated for discrete-variable (DV) optical systems~\cite{93Zukowki,98Pan,02Sciarrino,05Riedmatten}.
The protocol starts with two entangled pairs, $A$-$B$ and $C$-$D$, each represented either by twin photons or by a single photon split into two distinct modes [Fig.~\ref{NewFig1}(a)].
A joint projective measurement of $B$ and $C$ onto one of the four two-qubit Bell states then leads to an entangled state for $A$ and $D$, even though $A$ and $D$ never directly interact with each other. Entanglement swapping can also be interpreted as the transfer of one half of an entangled state, either from $B$ to $D$ or from $C$ to $A$, by quantum teleportation~\cite{93Bennett}.
It is a key element for quantum networking~\cite{98Bose}, quantum computing~\cite{01Knill},
quantum cryptography~\cite{12Braunstein},
and especially long-distance quantum communication by means of quantum repeaters~\cite{98Briegel,01Duan}. However, in this DV setting solely based upon single photons, due to the heralded conditional
entangled-state generation and the probabilistic linear-optics Bell-state measurement (BSM), successful entanglement swapping events occur very rarely. As a result, in a quantum repeater for example, long-distance entangled-pair creation rates would be correspondingly low and requirements on the coherence times of the local quantum memories at each repeater station impractically high. In addition, the observation and verification of the final entanglement between $A$ and $D$ in the DV scheme typically requires post-selection.

Entanglement swapping was later extended to continuous-variable (CV) systems \cite{12Weedbrook,05Braunstein},
where the pairs $A$-$B$ and $C$-$D$ each correspond to the two modes of a two-mode squeezed, quadrature-entangled state~\cite{99Tan,99vanLoock} [Fig.~\ref{NewFig1}(b)]. Since such entangled states are available on demand and a linear-optics BSM in the quadrature basis can be performed without failure, entanglement can be swapped
deterministically and verified without post-selection~\cite{04Jia,05Takei}. However, due to the finite squeezing of both initial entanglement sources, the final entanglement after swapping in the CV scheme is inevitably degraded by excess noise. In fact, the entanglement drops exponentially~\cite{PvLFdP,Ohliger} and in practice, CV entanglement swapping can always be replaced by a direct transmission through a lossy channel~\cite{Hoelscher}. Moreover, purification techniques for this type of degraded entanglement are not so advanced at present~\cite{10Takahashi,10Xiang}.

Our scheme combines the best features of the above two approaches, making use of both DV and CV entanglement at the same time [Fig.~\ref{NewFig1}(c)]. By means of CV teleportation~\cite{98Furusawa,13Takeda2},
using squeezed-state entanglement, homodyne-based BSM, and feedforward by phase-space displacement, DV entanglement is transferred from $A$-$B$ to $A$-$D$. Once the initial single-photon entanglement is conditionally prepared in modes $A$ and $B$, all the remaining steps of our scheme are unconditional, achieving a highly efficient transfer of the DV entanglement. Because optical CV quantum teleportation runs in a deterministic fashion (as opposed to optical DV quantum teleportation) and DV entanglement is robust against loss (as opposed to CV entanglement), entanglement is efficiently and reliably transferred only in this ``hybrid'' setting.
Furthermore, as will be explained below, a maximally entangled state can be obtained after the swapping through post-selection,
even though only finitely squeezed resources are used.

\begin{figure}[!t]
  \begin{center}
   \includegraphics[width=\linewidth,clip]{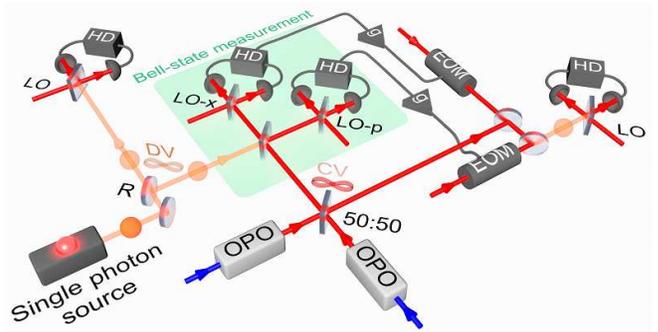}
  \end{center}
\caption{Experimental layout.
DV entanglement is created by splitting a heralded single photon~\cite{13Takeda3},
while CV entanglement is created on demand by mixing
two squeezed beams from optical parametric oscillators (OPOs).
The state after entanglement swapping is characterized by homodyne tomography~\cite{04Babichev,04Lvovsky}.
EOM; electro-optic modulator, HD; homodyne detector, and LO; local oscillator.}
\label{NewFig2}
\vspace{-5mm}
\end{figure}

In this hybrid setting, the DV entanglement can be transferred for any nonzero squeezing, as is
theoretically shown in Refs.~\cite{99Polkinghorne,13Takeda1}. Our setup (Fig.~\ref{NewFig2}) uses the
DV entanglement in the form of a photon split at a beam splitter with reflectivity $R$, described by
$\ket{\psi}_{AB}=\sqrt{1-R}\ket{1}_A\ket{0}_B+\sqrt{R}\ket{0}_A\ket{1}_B$ in the two-mode photon number basis.
This state is maximally entangled when $R=0.5$. In contrast, the CV entanglement is a two-mode squeezed state, $\sqrt{1-g_\text{opt}}\sum_{n=0}^{\infty}g_\text{opt}^n\ket{n}_C\ket{n}_D$ with $g_\text{opt}\equiv\tanh r$, where $r$ is the squeezing parameter. Though this state is nonmaximally entangled for finite $r$, the DV entanglement can be transferred for any $r>0$ by tuning the feedforward gain to $g_\text{opt}$, when the final state of $D$ is an imperfect version of the initial state of $B$ attenuated by a factor $1-g_\text{opt}^2$~\cite{99Polkinghorne,13Takeda1}. At this gain, the initial entangled state at $R=0.5$ is swapped and transformed according to
\begin{align}
\hat{\rho}_{AB}&\!\equiv\!\ket{\psi}_{AB}\!\bra{\psi} \nonumber\\
\to\hat{\rho}_{AD}&\!\equiv\!\frac{1\!+\!g_\text{opt}^2}{2}\!\ket{\psi^\prime}_{AD}\!\bra{\psi^\prime}\!+\!\frac{1\!-\!g_\text{opt}^2}{2}\!\ket{0,0}_{AD}\!\bra{0,0},
\label{eq:optimal_swapping}
\end{align}
where
$\ket{\psi^\prime}_{AD}=(\ket{1}_A\ket{0}_D+g_\text{opt}\ket{0}_A\ket{1}_D)/(1+g_\text{opt}^2)^{1/2}$. Here, the
initial maximally entangled state $\ket{\psi}_{AB}$ is converted into a nonmaximally entangled state
$\ket{\psi^\prime}_{AD}$ mixed with an extra two-mode vacuum term.
When $g_\text{opt}>0$, $\hat{\rho}_{AD}$ violates the positivity after partial transposition;
this shows that DV entanglement remains present after teleportation for any $r>0$ by optimal gain tuning~\cite{96Peres}.
Since no additional photons are created in $\hat{\rho}_{AD}$, it can be used for teleportation~\cite{02Lombardi}, swapping~\cite{02Sciarrino}, and purification
protocols~\cite{10Salart}.

The present experimental setup (Fig.~\ref{NewFig2}) is an extended version of the setup in \cite{13Takeda2}. We use a continuous-wave Ti:sapphire laser at 860 nm. A heralded single photon with a HWHM of 6.2 MHz is created from a nondegenerate optical parametric oscillator (OPO) at a rate of 7000 s$^{-1}$~\cite{13Takeda3}. The photon when incident on a beam splitter of reflectivity $R=0.50$ or $R=0.67$ yields a DV entangled state $\ket{\psi}_{AB}$. The CV entangled state is deterministically generated by combining at a beam splitter two squeezed vacua each produced from a degenerate OPO with a HWHM of 12 MHz. A CV BSM is performed jointly on the two corresponding halves of these two entangled states by combining them at a 50:50 beam splitter and then measuring the
orthogonal quadratures of the output modes by homodyne detection.
The homodyne signals are multiplied by a factor $g$ and used for modulating auxiliary coherent beams.
These beams are combined with the other half of the CV entangled state, thereby displacing the state in phase space.
Tomographic reconstruction of the initial and final states, $\hat{\rho}_{AB}$ and $\hat{\rho}_{AD}$,
are performed by two homodyne measurements with local oscillators' phases $\theta_1$ and $\theta_2$.
For these particular states, the sum $\theta_1+\theta_2$ does not affect the homodyne statistics in theory~\cite{13Takeda1}.
Thus, we first confirm the sum independence of the homodyne statistics and then
scan only the relative phase $\theta_1-\theta_2$ for tomography~\cite{04Babichev}.
For every state, 100,000 sets of quadrature and phase values are acquired and used for
a maximum likelihood algorithm without compensation of the measurement inefficiency~\cite{04Lvovsky}.

\begin{figure*}[!t]
  \begin{center}
   \includegraphics[width=0.9\linewidth,clip]{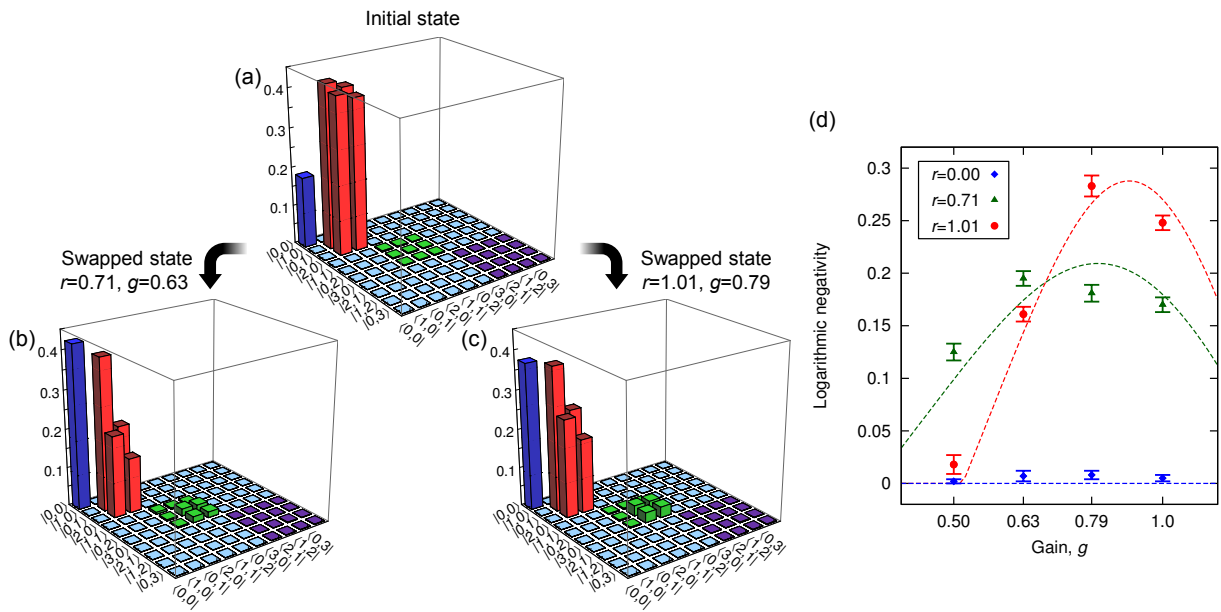}
  \end{center}
\caption{Experimental results.
(a)-(c) Density matrices of $\hat{\rho}_{AB}$ [(a)], $\hat{\rho}_{AD}$ at $r=0.71$
and $g=0.63$ [(b)], and $\hat{\rho}_{AD}$ at $r=1.01$ and $g=0.79$ [(c)]. The absolute value of each matrix element
is plotted. (d) Gain dependence of the logarithmic negativity for $r=0$ (blue diamonds), $r=0.71$ (green triangles), and
$r=1.01$ (red circles). Theoretical curves are also plotted in the same colors.}
\label{NewFig3}
\vspace{-5mm}
\end{figure*}

The experimental density matrix of the initial DV entangled state, $\hat{\rho}_{AB}$, at $R=0.5$ is
shown in Fig.~\ref{NewFig3}(a). This state includes $80.6\pm0.3\%$ of the ideal $\ket{\psi}_{AB}$, $18.3\pm0.3\%$ of vacuum, and $1.1\pm0.2\%$ of multiphoton terms. The density matrices of the swapped states, $\hat{\rho}_{AD}$, at $r=0.71$, $g=0.63$ ($g_\text{opt}=0.61$) and $r=1.01$, $g=0.79$ ($g_\text{opt}=0.77$) are also shown in Figs.~\ref{NewFig3}(b) and (c), respectively~\cite{Note}. It can be seen that only one mode of the entangled state is attenuated by a factor close to $1-g_\text{opt}^2$, but the off-diagonal elements ($\ket{0,1}\!\bra{1,0}$ and $\ket{1,0}\!\bra{0,1}$) are still preserved, indicating that the DV entanglement remains present after the swapping. The amount of entanglement can be assessed by the logarithmic negativity, $E( \hat \rho )\equiv\log_2 || \hat \rho^{\Gamma} ||$, where $||\hat \rho || \equiv \text{Tr}(\hat \rho ^\dagger \hat \rho)^{1/2}$ is the trace norm and $\Gamma$ denotes partial transposition with regards to one of the subsystems \cite{02Vidal,05Plenio,01Eisert}.
The gain dependence of $E( \hat \rho_{AD} )$ at $r=0$, $r=0.71$ and $r=1.01$ is plotted in Fig.~\ref{NewFig3}(d). Though the values of $E( \hat \rho_{AD} )$ are reduced from the initial value of $E( \hat \rho_{AB} )=0.71\pm0.01$, the positive values of $E( \hat \rho_{AD} )$ at $r>0$ clearly demonstrate the successful entanglement swapping. As expected, no entanglement is observed for $r=0$. These results also confirm the quantum nature of CV quantum teleportation~\cite{98Furusawa} in transferring DV systems~\cite{13Takeda2}. The transferred entanglement from $A$-$B$ to $A$-$D$
[$E(\hat{\rho}_{AD})=0.28\pm0.01$ at the maximum] is much greater than in the previous swapping experiments for
discrete variables~\cite{98Pan,02Sciarrino,05Riedmatten}, which post-selectively transferred the initial entanglement with a probability less than $1\%$, corresponding to $E(\hat{\rho}_{AD})<0.01$ without post-selection. We also performed the experiment for $R=0.67$ and observed $E( \hat \rho_{AD} )>0$ for $r=0.71$ and $r=1.01$~\cite{SOM}.

The final state, $\hat{\rho}_{AD}$, is contaminated with an extra vacuum due to the finite squeezing $r$;
however, it can be, in principle, purified to a maximally
entangled state post-selectively~\cite{01Duan}.
Remarkably, this also works for any $r>0$, even though the pure-state component in $\hat{\rho}_{AD}$ is itself only a nonmaximally entangled state for finite squeezing (as opposed to the maximally entangled Bell-state fractions in the scheme of Ref.~\cite{01Duan}).
The purification is achieved by first preparing two copies of
$\hat{\rho}_{AD}$, written as $\hat{\rho}_{A_1D_1}\otimes\hat{\rho}_{A_2D_2}$, and then projecting this state onto the
subspace with one photon in each location, corresponding to
$\{\ket{1}_{A_1}\ket{0}_{A_2},\ket{0}_{A_1}\ket{1}_{A_2}\}\otimes
\{\ket{1}_{D_1}\ket{0}_{D_2},\ket{0}_{D_1}\ket{1}_{D_2}\} \equiv \{\ket{A_1},\ket{A_2}\}\otimes
\{\ket{D_1},\ket{D_2}\}$. When $\hat{\rho}_{AD}$ has the form of Eq.~(\ref{eq:optimal_swapping}), this projection leads to a maximally entangled state, $(\ket{A_1}\ket{D_2}+\ket{A_2}\ket{D_1})/\sqrt2$, regardless of the squeezing level
$r>0$.
In other words, in principle, the hybrid setting allows for transferring maximally entangled states by means of finitely squeezed resources 
with a finite success probability, which in our experiment is already an order of magnitude larger compared to Refs.~\cite{98Pan,02Sciarrino,05Riedmatten}
(see below) and can be further increased for higher squeezing.

\begin{figure}[!t]
  \begin{center}
   \includegraphics[width=0.8\linewidth,clip]{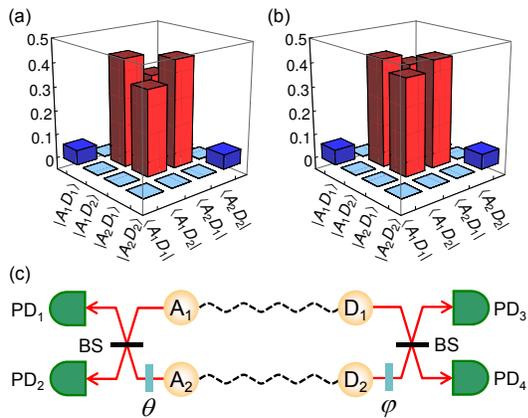}
  \end{center}
\caption{Analysis with post-selection.
(a) $\hat{\rho}_{AD}^\text{ps}$ calculated from $\hat{\rho}_{AD}$ at
$r=0.71$ and $g=0.63$ [Fig.~\ref{NewFig3}(b)].
(b) $\hat{\rho}_{AD}^\text{ps}$ calculated from $\hat{\rho}_{AD}$
at $r=1.01$ and $g=0.79$ [Fig.~\ref{NewFig3}(c)].
(c) Schematic setup for the realization of Bell's inequality
detection with post-selection. The values of the $S$ parameter estimated in the main text correspond to those values that would be measured in this setup. The setup consists of two entangled pairs ($A_1$-$D_1$, $A_2$-$D_2$), phase shifters ($\theta$,
$\phi$), 50:50 beam splitters (BS), and photon detections (PD$_1$-PD$_4$).}
\label{NewFig4}
\vspace{-5mm}
\end{figure}

We perform this purification protocol by analytically extracting the corresponding subspace from two copies of
the experimental $\hat{\rho}_{AD}$~\cite{SOM}. The renormalized density matrices after post-selection, $\hat{\rho}_{AD}^\text{ps}$, calculated from $\hat{\rho}_{AD}$ in Figs.~\ref{NewFig3}(b) and (c), are shown in Figs.~\ref{NewFig4}(a) and (b), respectively.
The probability for the state being projected onto $\hat{\rho}_{AD}^\text{ps}$ is $12.5\pm0.2\%$ and
$16.0\pm0.3\%$, respectively, which is calculated as the trace of the post-selected subspace. It can be seen that both
states are almost purified to the maximally entangled state $(\ket{A_1}\ket{D_2}+\ket{A_2}\ket{D_1})/\sqrt2$. For the
ideal $\hat{\rho}_{AD}$ in Eq.~(\ref{eq:optimal_swapping}), the $\ket{A_1D_1}\bra{A_1D_1}$ and $\ket{A_2D_2}\bra{A_2D_2}$ elements should be zero after post-selection. The small contributions of these terms in
Figs.~\ref{NewFig4}(a) and (b) originate from the multiphoton term $\ket{1,1}\bra{1,1}$ in $\hat\rho_{AD}$, and this term is mainly attributed to the impurity of squeezing. The values of the logarithmic negativity,
$E(\hat{\rho}_{AD}^\text{ps})=0.67\pm0.02$ for Fig.~\ref{NewFig4}(a) and
$E(\hat{\rho}_{AD}^\text{ps})=0.75\pm0.02$ for Fig.~\ref{NewFig4}(b), are greater than those without
post-selection, demonstrating the purification of the entanglement. In addition, the post-selected state can be used
for measuring violations of Bell's inequality by the setup shown in Fig.~\ref{NewFig4}(c)~\cite{01Duan}. Our calculation shows that the post-selected state in Figs.~\ref{NewFig4}(a) and (b) can, in principle, violate
the Clauser-Horne-Shimony-Holt inequality~\cite{69Clauser} by the estimated $S$ parameters of $S=2.08\pm0.05>2$ and $S=2.21\pm0.05>2$, respectively
(the latter indicates the violation by more than four standard deviations).
These highly entangled states can be directly used for quantum key distribution with the same setup as in Fig.~\ref{NewFig4}(c)~\cite{01Duan}.
These states would also enable one to do quantum teleportation of qubits~\cite{SOM},
and the estimated fidelities of $0.86\pm0.01$ and $0.89\pm0.01$ are well beyond the classical limit~\cite{95Massar} of $2/3$.

In conclusion, we demonstrated a ``hybrid'' entanglement swapping scheme, transferring robust DV entanglement in the form of a split photon by means of efficient CV entanglement and operations.
By tuning the feedforward gain of the teleporter, entanglement is reliably and efficiently transferred, and then verified unconditionally.
Moreover, despite the finite squeezing of the CV entanglement resource,
the DV states after the swapping can always be post-selected into highly entangled states that violate Bell's inequality
and may serve as resources for advanced quantum information protocols.
These results imply many possibilities for near-future applications of hybrid quantum networks, where
more general forms of DV entanglement may be efficiently transferred or manipulated with the help of CV techniques.
In a realistic network, where both the DV and the CV parts would be subject to transmission losses during their distributions, it is then better
to have the loss-sensitive CV links shorter than the loss-robust DV links or, in the most extreme scenario, to employ the CV entanglement only as 
a local on-demand resource, for instance, in order to deterministically load some local quantum memories~\cite{13Yoshikawa}.

\begin{acknowledgments}
This work was partially supported by the SCOPE program of the MIC of Japan, PDIS, GIA, and APSA commissioned by the
MEXT of Japan, FIRST initiated by CSTP, and ASCR-JSPS. S.T. and M.F. acknowledges financial support from ALPS.
P.v.L. was supported by QcomQ (BMBF) and HIPERCOM (ERA-Net CHIST-ERA).
We thank Masato Koashi for useful discussions.
\end{acknowledgments}


\clearpage

\begin{center}
\bf
\large{Supplemental material:\\ Entanglement swapping between discrete and continuous variables}
\end{center}

\section{Purification by post-selection}

Here we will show how the entangled state after the entanglement swapping,
\begin{align}
\hat{\rho}_{AD}=\sum_{k,l,m,n}\rho_{klmn}\ket{k}_A\ket{l}_{D\>A}\!\bra{m}_{\>D}\!\bra{n}\,,
\label{eq:rho_original}
\end{align}
written in the photon number basis,
is purified by the post-selection method given in Ref.~\cite{S01Duan}.
First we consider the $\{\ket{0}_A\ket{0}_D,\ket{0}_A\ket{1}_D,\ket{1}_A\ket{0}_D,\ket{1}_A\ket{1}_D\}$ subspace
of Eq.~(\ref{eq:rho_original}), which has the matrix form~\cite{S13Takeda}
\begin{align}
\left[\begin{array}{cccc}
p_{00} & 0 & 0 & 0 \\
0 & p_{01} & d & 0 \\
0 & d^* & p_{10} & 0 \\
0 & 0 & 0 & p_{11} \end{array}\right],
\label{eq:rho_AD}
\end{align}
where $p_{ij}\ge0$ ($i,j=0,1$), $\sum_{i,j} p_{ij}\le1$, and $|d|^2\le p_{01}p_{10}$.
The quantity $p_{ij}$ is the probability of finding $i$ photons in mode $A$ and $j$ photons in mode $D$,
and the off-diagonal element $d$ represents the coherence between $\ket{0}_A\ket{1}_D$ and $\ket{1}_A\ket{0}_D$.
Next we assume that two copies of the entangled state of Eq.~(\ref{eq:rho_original}) are distributed between two locations
through the same swapping channels.
This state can be written as $\hat{\rho}_{A_1D_1}\otimes\hat{\rho}_{A_2D_2}$,
from which we post-select the cases when
there is only one photon in each location.
This operation extracts the subspace spanned by
$\{\ket{1}_{A_1}\ket{0}_{A_2}, \ket{0}_{A_1}\ket{1}_{A_2}\}\otimes
\{\ket{1}_{D_1}\ket{0}_{D_2}, \ket{0}_{D_1}\ket{1}_{D_2}\}
\equiv
\{\ket{A_1},\ket{A_2}\}\otimes
\{\ket{D_1},\ket{D_2}\}$ from the full density matrix $\hat{\rho}_{A_1D_1}\otimes\hat{\rho}_{A_2D_2}$.
The corresponding subspace can be written as
\begin{align}
\left[\begin{array}{cccc}
p_{00}p_{11} & 0 & 0 & 0 \\
0 & p_{01}p_{10} & |d|^2 & 0 \\
0 & |d|^2 & p_{01}p_{10} & 0 \\
0 & 0 & 0 & p_{00}p_{11} \end{array}\right],
\label{eq:rho_AD_ps}
\end{align}
in the basis $\{\ket{A_1D_1}, \ket{A_1D_2}, \ket{A_2D_1}, \ket{A_2D_2}\}$.
The density matrix of the post-selected state, $\hat{\rho}_{AD}^\text{ps}$, is obtained
by renormalizing the matrix of Eq.~(\ref{eq:rho_AD_ps}).
The probability that the initial state $\hat{\rho}_{A_1D_1}\otimes\hat{\rho}_{A_2D_2}$
is projected onto $\hat{\rho}_{AD}^\text{ps}$
is given by the trace of Eq.~(\ref{eq:rho_AD_ps})
as $P=2(p_{00}p_{11}+p_{01}p_{10})$.
When there is no $\ket{1,1}$ contribution in
$\hat{\rho}_{A_1D_1}$ and $\hat{\rho}_{A_2D_2}$ ($p_{11}=0$),
the $p_{00}p_{11}$ terms in Eq.~(\ref{eq:rho_AD_ps}) vanish.
In this case, only the original elements in the $\{\ket{0}_A\ket{1}_D,\ket{1}_A\ket{0}_D\}$ subspace in Eq.~(\ref{eq:rho_AD})
are post-selected and constitute the elements in the $\{\ket{A_1D_2}, \ket{A_2D_1}\}$ subspace in Eq.~(\ref{eq:rho_AD_ps}).
Especially, if $p_{11}=0$ and $|d|^2= p_{01}p_{10}$,
Eq.~(\ref{eq:rho_AD_ps}) becomes a pure maximally entangled state,
$(\ket{A_1D_2}+\ket{A_2D_1})/\sqrt2$, even when $p_{00}>0$ and $p_{01}\neq p_{10}$.
This means that the initial mixture of a \textit{non-maximally} entangled state \textit{with} the vaccum
is automatically purified through this post-selection to a pure, \textit{maximally} entangled state \textit{without} the vacuum.
In other words, the entanglement is concentrated and purified at the same time. 
For example, if $\hat{\rho}_{AD}$ is written as Eq.~(1) of the main text,
the state can be, in principle, purified to a maximally entangled state.

\section{Bell's inequality detection with post-selection}

\begin{figure*}[!ht]
\begin{center}
\includegraphics[scale=0.92,clip]{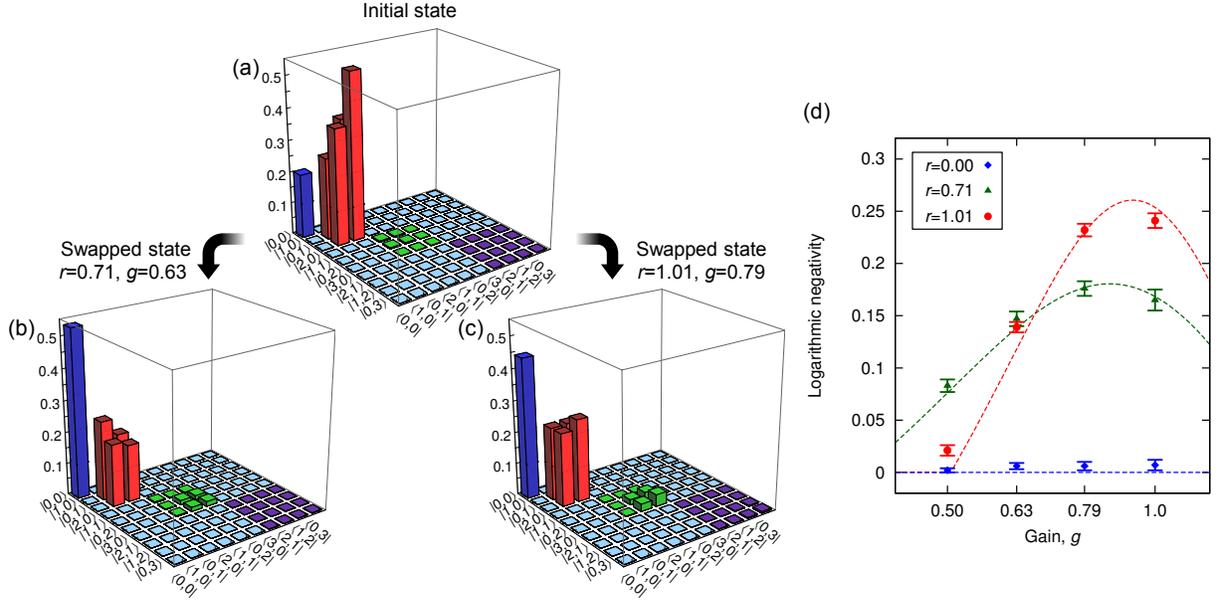}
\end{center}
\vspace{-5mm}
\caption{
Experimental results for $R=0.67$.
(a)-(c) Density matrices of
$\hat{\rho}_{AB}$ [(a)],
$\hat{\rho}_{AD}$ at $r=0.71$ and $g=0.63$ [(b)],
and $\hat{\rho}_{AD}$ at $r=1.01$ and $g=0.79$ [(c)].
The absolute value of each matrix element is plotted.
(d) Gain-dependence of the logarithmic negativity for $r=0$ (blue diamond), $r=0.71$ (green triangle), and $r=1.01$ (red circle).
Theoretical curves are also plotted in the same colors.
}
\label{fig:result_sup}
\vspace{-2mm}
\end{figure*}

Next we derive the condition when Bell's inequality is violated post-selectively with the setup of Fig.~4(c) in the main text.
Suppose that the coincidences of photon detectors are registered only when one photon is detected at both sides.
In this case, we only have to consider the post-selected elements of Eq.~(\ref{eq:rho_AD_ps}),
which is renormalized to
\begin{align}
\rho_{AD}^\text{ps}=
\left[\begin{array}{cccc}
(1-x)/2 & 0 & 0 & 0 \\
0 & x/2 & y/2 & 0 \\
0 & y/2 & x/2 & 0 \\
0 & 0 & 0 & (1-x)/2 \end{array}\right],
\label{eq:rho_AD_ps_renorm}
\end{align}
where
\begin{align}
x=2p_{01}p_{10}/P,\quad y=2|d|^2/P.
\label{eq:definition_xy}
\end{align}
An appropriate setting of the beam splitters and phase shifters in Fig.~4(c)
enables one to measure $\hat\rho_{AD}^\text{ps}$
in the rotated basis
$\{\cos\theta_A\ket{A_1}+\sin\theta_A\ket{A_2}, -\sin\theta_A\ket{A_1}+\cos\theta_A\ket{A_2}\}\otimes
\{\cos\theta_D\ket{D_1}+\sin\theta_D\ket{D_2}, -\sin\theta_D\ket{D_1}+\cos\theta_D\ket{D_2}\}$.
This measurement is equivalent to the measurement of
$\hat{R}^\dagger(\theta_A,\theta_D)\hat{\rho}_{AD}^\text{ps}\hat{R}(\theta_A,\theta_D)$
in the basis of $\{\ket{A_1},\ket{A_2}\}\otimes\{\ket{D_1},\ket{D_2}\}$, where
\begin{align}
&R(\delta,\phi)=\nonumber\\
&\left[\begin{array}{cccc}
\cos\delta\cos\phi & \cos\delta\sin\phi & \sin\delta\cos\phi & \sin\delta\sin\phi \\
-\cos\delta\sin\phi & \cos\delta\cos\phi & -\sin\delta\sin\phi & \sin\delta\cos\phi \\
-\sin\delta\cos\phi & -\sin\delta\sin\phi & \cos\delta\cos\phi & \cos\delta\sin\phi \\
\sin\delta\sin\phi & -\sin\delta\cos\phi & -\cos\delta\sin\phi & \cos\delta\cos\phi \\
\end{array}\right].
\label{eq:rotation}
\end{align}
describes a $\delta$ rotation in the $\{\ket{A_1},\ket{A_2}\}$ basis
and a $\phi$ rotation in the $\{\ket{D_1},\ket{D_2}\}$ basis.
The probability of obtaining a coincidence
between the detectors for $\ket{A_i}$ and $\ket{D_j}$ ($i,j=1,2$)
is given by
\begin{align}
P_{ij}(\theta_A,\theta_D)=\braket{A_iD_j|\hat{R}^\dagger(\theta_A,\theta_D)\hat\rho_{AD}^\text{ps} \hat{R}(\theta_A,\theta_D)|A_iD_j}.
\label{eq:def_P}
\end{align}
By substituting Eqs.~(\ref{eq:rho_AD_ps_renorm}) and (\ref{eq:rotation}) into Eq.~(\ref{eq:def_P}), we obtain
\begin{align}
P_{11}(\theta_A,\theta_D)&=P_{22}(\theta_A,\theta_D)\nonumber\\
&=\frac{1}{8}\{2+(1-2x+y)\cos[2(\theta_A-\theta_D)]\nonumber\\
&\hspace{8mm}+(1-2x-y)\cos[2(\theta_A+\theta_D)]\},\\
P_{12}(\theta_A,\theta_D)&=P_{21}(\theta_A,\theta_D)\nonumber\\
&=\frac{1}{8}\{2-(1-2x+y)\cos[2(\theta_A-\theta_D)]\nonumber\\
&\hspace{8mm}-(1-2x-y)\cos[2(\theta_A+\theta_D)]\}.
\end{align}
For a detection of Bell's inequality,
we estimate a correlation function defined as
\begin{align}
&E(\theta_A,\theta_D)\nonumber\\
&=P_{11}(\theta_A,\theta_D)\!-\!P_{12}(\theta_A,\theta_D)\!-\!P_{21}(\theta_A,\theta_D)\!+\!P_{22}(\theta_A,\theta_D)\nonumber\\
&=\frac{1}{2}\{(1-2x+y)\cos[2(\theta_A-\theta_D)]\nonumber\\
&\hspace{25mm}+(1-2x-y)\cos[2(\theta_A+\theta_D)]\},
\label{eq:correlation_function}
\end{align}
and then assess the parameter
\begin{align}
S\!=\!\left|E(\theta_A,\theta_D)\!+\!E(\theta_A^\prime,\theta_D)\!-\!E(\theta_A,\theta_D^\prime)\!+\!E(\theta_A^\prime,\theta_D^\prime)\right|.
\label{eq:definition_S}
\end{align}
The Clauser-Horne-Shimony-Holt (CHSH) inequality implies that $S$ should be below 2 for any local hidden variable theories~\cite{S69Clauser}.
Here we set $(\theta_A,\theta_A^\prime,\theta_D,\theta_D^\prime)=(0,\pi/4,3\pi/8,\pi/8)$,
and then substitute Eq.~(\ref{eq:correlation_function}) into Eq.~(\ref{eq:definition_S})
to obtain
\begin{align}
S&=\left|\sqrt2(2x+y-1)\right|.
\label{eq:Sparameter}
\end{align}
This formula gives an estimated $S$ parameter for a given experimental $\hat{\rho}_{AD}$.
$S>2$ indicates that the state potentially violates the CHSH inequality post-selectively.
In the ideal case of $x=y=1$, Eq.~(\ref{eq:Sparameter}) gives $S=2\sqrt2>2$, and thus the inequality can be violated.

A straightforward way to experimentally implement this Bell's inequality detection is
to first prepare two pairs of DV and CV entangled states and then perform the entanglement swapping in parallel.
After the swapping, the initial state $\hat{\rho}_{A_1D_1}\otimes\hat{\rho}_{A_2D_2}$ is ready for the Bell's inequality detection.
The local rotation $R(\delta,\theta)$ of Eq.~(\ref{eq:rotation}) can be realized by combining $A_1$, $A_2$ and $D_1$, $D_2$ at 50:50 beam splitters
with their relative phases locked to $2\delta$ and $2\phi$ by the standard CV technique.
Finally each port of the beam splitters is subject to photon measurements,
and only the appropriate coincident events are post-selected to evaluate the $S$ parameter.

\section{Quantum teleportation with post-selection}

\begin{figure}[!t]
\begin{center}
\includegraphics[scale=0.6]{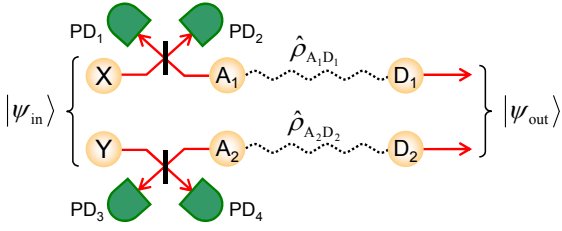}
\end{center}
\vspace{-5mm}
\caption{
Schematic of quantum teleportation of qubits with post-selection.
The input qubit encoded in mode $X$ and $Y$ is teleported to mode $D_1$ and $D_2$.
}
\label{fig:teleportation}
\vspace{-2mm}
\end{figure}

The shared entangled state $\hat{\rho}_{AD}$ of Eq.~(\ref{eq:rho_original}) can be used for quantum teleportation of qubits with post-selection~\cite{S01Duan}.
The schematic is shown in Fig.~\ref{fig:teleportation}.
Two entangled pairs, $\hat{\rho}_{A_1D_1}\otimes\hat{\rho}_{A_2D_2}$, are first prepared as a resource of teleportation.
An unknown qubit is encoded in mode $X$ and $Y$ as
\begin{align}
\ket{\psi_\text{in}}=\alpha\ket{1}_X\ket{0}_Y\!+\!\beta\ket{0}_X\ket{1}_Y\equiv \alpha\ket{X}\!+\!\beta\ket{Y},
\label{eq:input_qubit}
\end{align}
where $|\alpha|^2+|\beta|^2=1$.
Mode $X$, $A_1$ and $Y$, $A_2$ are then subjected to photon-counting measurement after 50:50 beam splitters.
If only PD1 and PD3 (or PD2 and PD4) detect single photons,
these modes are projected onto one of four Bell states, $\ket{\Psi}=(\ket{YA_1}+\ket{XA_2})/\sqrt2$.
After this projective measurement,
we post-select the case when the output modes $D_1$ and $D_2$ contain only one photon in total.
Ideally, a maximally entangled state $(\ket{A_1D_2}+\ket{A_2D_1})/\sqrt2$ is extracted
from $\hat{\rho}_{A_1D_1}\otimes\hat{\rho}_{A_2D_2}$ after this projection and post-selection.
The post-selected output state is thus written as
\begin{align}
\bra{\Psi}
\left(\ket{\psi_\text{in}}\otimes\frac{\ket{A_1D_2}+\ket{A_2D_1}}{\sqrt2}\right)
=\frac{1}{2}\left(\alpha\ket{D_1}+\beta\ket{D_2}\right),
\end{align}
which is normalized into $\ket{\psi_\text{out}}=\alpha\ket{D_1}+\beta\ket{D_2}$;
the same qubit as Eq.~(\ref{eq:input_qubit}) appears at the output without any local operation,
and the qubit teleportation is achieved with unit-fidelity.

More generally, all the relevant density matrix elements of Eq.~(\ref{eq:rho_original}) have to be
taken into account to calculate the output state and the corresponding fidelity.
The projective measurement and post-selection finally extract
the subspace spanned by $\{\ket{YA_1D_1}, \ket{YA_1D_2}, \ket{XA_2D_1}, \ket{XA_2D_2}\}$
from the overall initial density matrix $\ket{\psi_\text{in}}\!\bra{\psi_\text{in}}\otimes\hat{\rho}_{A_1D_1}\otimes\hat{\rho}_{A_2D_2}$.
From Eqs.~(\ref{eq:rho_AD_ps}) and (\ref{eq:input_qubit}),
the matrix elements included in this subspace are
\begin{align}
\hat{\rho}_\text{ext}=&|\beta|^2 p_{00}p_{11}\ket{YA_1D_1}\!\bra{YA_1D_1} \nonumber\\
&\quad+|\beta|^2 p_{01}p_{10}\ket{YA_1D_2}\!\bra{YA_1D_2} \nonumber\\
&\quad+\alpha^*\beta |d|^2 \ket{YA_1D_2}\!\bra{XA_2D_1} \nonumber\\
&\quad+\alpha\beta^* |d|^2 \ket{XA_2D_1}\!\bra{YA_1D_2} \nonumber\\
&\quad+|\alpha|^2 p_{01}p_{10}\ket{XA_2D_1}\!\bra{XA_2D_1} \nonumber\\
&\quad+|\alpha|^2 p_{00}p_{11}\ket{XA_2D_2}\!\bra{XA_2D_2}.
\end{align}
After the projective measurement onto $\ket{\Psi}=(\ket{YA_1}+\ket{XA_2})/\sqrt2$,
the output state to be post-selected is given by
\begin{align}
\braket{\Psi|\hat{\rho}_\text{ext}|\Psi}
&=\frac{|\alpha|^2 p_{01}p_{10}\!+\!|\beta|^2 p_{00}p_{11}}{2}\ket{D_1}\!\bra{D_1} \nonumber\\
&\quad+\frac{\alpha\beta^*|d|^2}{2}\ket{D_1}\!\bra{D_2}
+\frac{\alpha^*\beta|d|^2}{2}\ket{D_2}\!\bra{D_1} \nonumber\\
&\quad+\frac{|\alpha|^2 p_{00}p_{11}\!+\!|\beta|^2 p_{01}p_{10}}{2}\ket{D_2}\!\bra{D_2},
\label{eq:tele_after_projection}
\end{align}
which is normalized into
\begin{align}
\hat{\rho}_\text{out}=
&\left[|\alpha|^2 x +|\beta|^2(1-x)\right]\ket{D_1}\!\bra{D_1} \nonumber\\
&\quad+\alpha\beta^*y\ket{D_1}\!\bra{D_2}+\alpha^*\beta y\ket{D_2}\!\bra{D_1} \nonumber\\
&\quad+\left[|\alpha|^2 (1-x) +|\beta|^2x\right]\ket{D_2}\!\bra{D_2}.
\end{align}
Here the definition of $x$ and $y$ are given by Eq.~(\ref{eq:definition_xy}).
The probability of successful teleportation is given by the trace of Eq.~(\ref{eq:tele_after_projection})
as $P/4$, where $P$ is previously defined as the trace of Eq.~(\ref{eq:rho_AD_ps}).
The factor of $1/4$ comes from the fact that the projection onto only one of four Bell states is considered here.
The fidelity of teleportation can be calculated from the output state $\hat{\rho}_\text{out}$ and the ideal output qubit $\ket{\psi_\text{out}}=\alpha\ket{D_1}+\beta\ket{D_2}$ as
\begin{align}
F(\alpha,\beta)&=\braket{\psi_\text{out}|\hat{\rho}_\text{out}|\psi_\text{out}}\nonumber\\
&=(|\alpha|^4+|\beta|^4)x+2|\alpha|^2|\beta|^2(1-x+y).
\end{align}
This value depends on the coefficients $\alpha$ and $\beta$ of the input qubit.
The fidelity averaged over all qubits is thus given by
\begin{align}
F_\text{av}
&=\frac{1}{4\pi}\int_{\theta=0}^{\theta=\pi}\int_{\phi=0}^{\phi=2\pi}
F\left(\cos\frac{\theta}{2},e^{i\phi}\sin\frac{\theta}{2}\right)\sin\theta d\theta d\phi \nonumber\\
&=\frac{1+x+y}{3}.
\end{align}
Therefore the average fidelity beyond the classical limit~\cite{S95Massar} of 2/3
is achieved when $x+y>1$.

\section{Experimental results for $R=0.67$}

In our experiment,
the beam splitter reflectivity $R$
for generating the discrete-variable entangled state
$\ket{\psi}_{AB}=\sqrt{1-R}\ket{1}_A\ket{0}_B+\sqrt{R}\ket{0}_A\ket{1}_B$
is set to two values, $R=0.50$ and $R=0.67$.
Below we show the results for $R=0.67$, which are not shown in the main text.

The experimental density matrices of the initial ($\hat{\rho}_{AB}$)
and final ($\hat{\rho}_{AD}$)
state as well as the logarithmic negativity of the final state
are summarized in Fig.~\ref{fig:result_sup}.
It can be seen from Fig.~\ref{fig:result_sup}(a)-(c) that
one mode of the initial state is attenuated by a factor $1-g_\text{opt}^2$ through the entanglement swapping channel,
as expected from theory.
Positive values of the logarithmic negativity $E( \hat \rho_{AD} )$ are obtained for $r>0$, 
as can be seen in Fig.~\ref{fig:result_sup}(d),
which verifies the successful entanglement swapping without post-selection.
In the case of $R=0.67$,
the initial discrete-variable entanglement is an asymmetric, non-maximally entangled state,
and thus the logarithmic negativity of the initial state
$E( \hat \rho_{AB} )=0.64\pm0.01$ is slightly below the value for $R=0.50$.
As a result, the logarithmic negativity of the final state
is also below the value for $R=0.50$ at the same $r$ and $g$ values.

\begin{figure}[!ht]
\begin{center}
\includegraphics[scale=0.6]{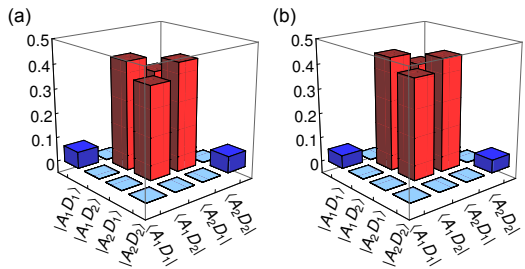}
\end{center}
\vspace{-5mm}
\caption{
Analysis with post-selection for $R=0.67$.
(a) $\hat{\rho}_{AD}^\text{ps}$ calculated from $\hat{\rho}_{AD}$ at $r=0.71$ and $g=0.63$ [Fig.~\ref{fig:result_sup}(b)].
(b) $\hat{\rho}_{AD}^\text{ps}$ calculated from $\hat{\rho}_{AD}$ at $r=1.01$ and $g=0.79$ [Fig.~\ref{fig:result_sup}(c)].
}
\label{fig:post_selection_sup}
\vspace{-2mm}
\end{figure}

We analytically perform the post-selection from $\hat{\rho}_{AD}$ in Figs.~\ref{fig:result_sup}(b) and (c).
Renormalized density matrices after the post-selection, $\hat{\rho}_{AD}^\text{ps}$,
are shown in Figs.~\ref{fig:post_selection_sup}(a) and (b).
The probability of the state being projected onto $\hat{\rho}_{AD}^\text{ps}$ is $10.3\pm0.2\%$ and $13.4\pm0.3\%$, respectively.
The values of the logarithmic negativity, $E(\hat{\rho}_{AD}^\text{ps})=0.70\pm0.04$ for Fig.~\ref{fig:post_selection_sup}(a) and
$E(\hat{\rho}_{AD}^\text{ps})=0.77\pm0.02$ for Fig.~\ref{fig:post_selection_sup}(b),
are greater than those without post-selection,
demonstrating the purification of the entanglement.
In addition, the post-selected state can
violate the CHSH inequality by
the estimated $S$ parameters of $S=2.11\pm0.08>2$ and $S=2.26\pm0.04>2$, respectively.
These states also enable quantum teleportation of qubits with the fidelities of $0.87\pm0.01>2/3$ and $0.90\pm0.01>2/3$, respectively.
These results and our analysis further support the demonstration of unconditional entanglement swapping
and the possibility of purification by post-selection.


\begin{thebibliography}{}
\bibitem{Note}
The experimental squeezed and antisqueezed quadrature variance is equivalent
to $[(1-l)e^{\pm 2r}+l]/2$ ($\hbar=1$),
where the estimated loss $l$ ranges from 0.20 to 0.24.
In the case of $l=0.22$, $r=0.71$ gives $-3.9$ dB of squeezing and $5.4$ dB of antisqueezing,
while $r=1.01$ gives $-4.9$ dB of squeezing and $7.9$ dB of antisqueezing.
\end{thebibliography}
\end{document}